\documentclass[a4paper,10pt,fleqn]{article}
\pdfoutput=1
\usepackage{jheppub}
\usepackage[T1]{fontenc}
\usepackage{setspace}
\usepackage{enumitem}
\usepackage{amsmath}
\usepackage{amsthm}
\usepackage{bm}
\usepackage{mathtools}
\usepackage{caption}
\usepackage{subcaption}
\usepackage{tikz}
\usepackage{float}
\usepackage{syllogism}
\usepackage{endnotes}

\let\footnote=\endnote

\usepgflibrary{arrows}
\usetikzlibrary{calc}

\renewcommand{\vec}[1]{\mathbf{#1}}

\renewcommand{\>}{\rangle}
\newcommand{\<}{\langle}

\renewcommand{\vec}[0]{\boldsymbol}

\title{The Ontology of Compositeness Within Quantum Field Theory}

\author[a]{T.~Peterken}

\affiliation[a]{School of Physics, University of Edinburgh, Edinburgh EH9 3JZ, UK}

\emailAdd{t.peterken@sms.ed.ac.uk}

\abstract{
In this work, we attempt to define a notion of compositeness compatible with Quantum Field Theory.  Considering the analytic properties of the S-matrix, we conclude that there is no satisfactory definition of compositeness compatible with Quantum Field Theory.  Without this notion, one must claim that all bound states are equally fundamental, that is, one cannot rigorously claim that everyday objects are made of atoms or that atoms are made of protons and neutrons.  I then show how an approximate notion of compositeness may be recovered in the regime where the mass of a bound state is close to a multi-particle threshold.

Finally, we see that rejecting compositeness solves several of the "problems of everyday objects" encountered in an undergraduate metaphysics course.
}

\begin{document} 

\maketitle
\flushbottom
\abovedisplayskip 11pt
\belowdisplayskip 11pt
\section{Introduction}

I began questioning the nature of compositeness within Quantum Field Theory (QFT) when writing my first literature review.  When talking about different particles, papers classified them as conventional hadrons, exotic hadrons, hadronic molecules, and so on. I could not find a convincing explanation of the difference between these categories of particles.

As bound states manifest themselves as poles in a scattering amplitude \cite{duncan2012conceptual}, their properties (such as mass, width, or even just their existence) cannot be calculated using a perturbative framework.  An alternative approach would be to use lattice field theory for these calculations (these techniques are covered in many books such as \cite{Gattringer} or in relevant review articles such as \cite{padmanath2019hadron}).

As an example, (which is only chosen as it is related to what I was reading at the time): in lattice field theory calculations the existence and mass of a stable bound state can be found from an appropriate time-dependent (Euclidean) correlator.  A bound state of mass $M$ causes the correlator to exhibit the following time dependence, up to discretization and volume effects: 
\begin{align}
    C(t,\vec{P})=\<\sigma(t,\vec{P})\sigma^\dagger\>\propto e^{-t\sqrt{M^2+\vec{P^2}}}+\text{ scattering states}
\end{align}
where $\sigma$ is an interpolation operator with a specified set of quantum numbers of the single-particle state. 

The details are unimportant, except that all stable bound states show the same behaviour, independent of if they would be considered conventional hadrons or not.  At the time, it felt as if these methods obscured the  difference between the different categories of particles - in hindsight, I would say it emphasized the similarities.

Eventually, I came to the conclusion that the different types of bound states are actually just a form of cataloguing: conventional hadrons are those whose quantum numbers can be produced from a simple combination of valence quarks whereas hadron-hadron molecules have the quantum numbers of a two-hadron channel, and have a mass that is just below this two-particle threshold \cite{donoghue_golowich_holstein_2014,Guo_2018}.  However, cataloguing based on phenomenological characteristics does not always give a good picture of the underlying reality.  

Biology is full of different categories, but there are always exceptions and ambiguous cases (as exemplified by the name of the podcast ``No Such Thing as a Fish" \cite{nosuchthingasafish} or that at first people didn't believe in the existence of a duck-billed platypus due to it not fitting into the pre-established categories of animals \cite{ohlheiser_2021} - despite the overwhelming evidence of 129 Phineas and Ferb episodes \cite{wikipedia_2023}).  Categorization is useful for sorting our observations, but not so useful for providing a fundamental understanding.  

In this work, I explore the notion of compositeness and its relation to QFT, and come to the potentially unsavoury conclusion that \textit{no satisfactory, exact, and rigorous definition of compositeness exists that is compatible with QFT}.  I start by justifying why it is reasonable to take the unobservable aspects of a physical theory seriously.  I then go on to elucidate what I require from a satisfactory definition of compositeness.  I then present two arguments as to why this satisfactory definition is not compatible with QFT and explain a few corollaries of this result, the first argument is best considered as a warm-up using perturbation theory, and the second argument as the main result of this work.  I then go through a range of possible objections and show that they don't save the notion of compositeness.  I then show how it is possible to define an approximate notion of compositeness and how it can be included in the ontology of higher-level theories such as molecular or solid-state physics.  Finally, I relate this to the problem of ordinary objects~\cite{sep-ordinary-objects} as encountered in introductory metaphysics courses and show that many of these problems dissolve if compositeness is rejected.

I assume the audience of this work is familiar with QFT, at the level of a graduate course, with some knowledge of its applications to particle physics and some of the formal aspects of scattering theory such as the LSZ procedure and the analytic structure of the S-matrix (chapter 7 of ref.~\cite{Peskin:1995ev} should suffice).

\section{Why should we take the unobservable seriously?}
Questions about interpreting scientific theories, or about the ontological status of certain aspects of a scientific theory (that is, questions about if and how the features of a scientific theory exist) often seem to be ignored by working physicists - at least in their professional work.  In fact, personally, I have found many people to dismiss such questions as `too philosophical' and not really worth thinking about.  In this section, I explain why I feel this dismissal is often too premature.

 QFT is hopefully not a theory that merely relates the center of mass energy of a hadron collider to some numbers on the screen; hopefully it is not just a theory relating free particles from infinitely in the past to free particles infinitely in the future - even if this is what is directly observable and well-defined via the LSZ procedure.  Something happens, in the real physical world, between starting an experiment at the Large Hadron Collider (LHC) and seeing numbers on a screen (or rather, something is going on between the asymptotic past and future states).  
 
 We may never be able to directly observe what is going on, we may never be certain about what is going on, we may never have a unique theory\footnote{Here I use the word `unique' differently to physicists. In physics, two theories are considered equivalent if they always give the same observational outcomes, but here I take a much stronger definition, two theories are unique if all content (observable and unobservable) is the same.  In this sense, a single mathematical formalism can give rise to several distinct theories depending on how it is interpreted.  More details on this are given in chapter 2 of ref.~\cite{interpretingQM}.} to tell us what is going on.  But \textit{something} \textbf{is} going on.  It would be remarkable if a theory that gave such accurate observational predictions was also completely wrong about everything else.  A  general introduction to the questions of scientific realism can be found in ref.~\cite{sep-scientific-realism, rosenberg2000philosophy}.

 There is a nice analogy (taken directly from the introduction of ref.~\cite{FAPP}) that highlights that this dismissal can often happen inconsistently, with people much less willing to take the realism of quantum mechanics as seriously as other theories.
 
 When we look at distant galaxies - so distant we will probably never get to them and the only thing we can do is look from afar -  all we can see is a sort of hazy glow. Using our understanding of galactic structure, we can infer that these galaxies are made from hundreds of billions of stars, that these stars will have planets around them, and that some of these planets will have atmospheres.  Even though we will never see these planets, I believe they exist and I believe they do have atmospheres, and I have not yet met a physicist who would claim to doubt this either.

 We believe in these atmospheres, on the ground that they are inferred from taking our best scientific theory seriously.  Why are they given a privilege when quantum fields are not? Even though the unobservability of extra-galactic atmospheres is due to practical limitations, no one will observe them in my lifetime and so when making the individual assessment of their existence, I can still only infer from the current best theory.  I now do the same with QFT.
\section{What is a satisfactory definition of compositeness?}
\label{sec:satisfactory}
I take the following to be necessary requirements for a satisfactory definition of compositeness.  These requirements are asserted in a loose manner as these are taken to be minimal requirements and having an overly precise definition would lead to this work rejecting a definition of compositeness that is too specific.

\begin{itemize}
    \item \textbf{For $X$ to be a composite object made of $A$ and $B$ we need to be able to refer to $A$ and $B$, in a well-defined way, whilst $X$ is in existence.}
    
    For example, a research group is made out of a collection of people, I can refer to both the research group and the members of the research group in a completely unambiguous way.  The members do not define the group, specific individuals can join and leave the group; however the group, at any given time, contains a set of individuals which can unambiguously be referred to.

    In non-relativistic quantum mechanics, the hydrogen atom is a composite object containing an electron and a proton.  We can unambiguously refer to the coordinates of the electron and proton separately to the whole atom. 
    
    \item \textbf{If $A$ is made of $B$, then $B$ is not made of $A$   -  that is, compositeness is not reflexive\footnote{Within this I also exclude the possibility of a circle of compositeness.  We cannot have $A$ containing $B$, $B$ containing $C$ and then $C$ containing A.}.}
    
   If an atom is made of a proton and an electron, then it would be absurd to say that the proton contained an atom as then an atom would contain a proton which would contain an atom ad infinitum.

    If you have a desire to save the notion of compositeness due to the historically successful explanatory power of reductionism, then this axiom is needed to save reductionist explanations from circularity.  
    
    \item \textbf{Reality cannot depend on arbitrary choices.}  
    
    In QFT, there are a huge range of choices I could make: gauge, renormalization scheme, I can rewrite the Lagrangian in a range of different ways, I could split up the Hamiltonian into a `free' and `interacting' part in a range of different ways and so on.  If I want to make a meaningful statement about the external world, then that statement can't depend on any of these arbitrary choices made.
    \end{itemize}

\section{The incompatibility of compositeness and QFT}
In this section we justify the following statement:
\begin{center}
    \textit{No satisfactory, exact (in the sense of being non-perturbative) and rigorous definition of compositeness exists compatible with QFT.}
\end{center}

\subsection{A warm-up from perturbation theory}
To explore compositeness exactly, any core argument cannot rest upon perturbation theory.  That being said, it lays the groundwork for most working physicists and we will see that hidden in perturbative calculations was a prophecy of the argument in the next section.

\begin{figure}
\centering
   \includegraphics[scale=1]{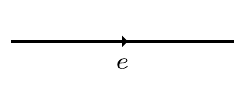}
   \caption{At leading order, the fully dressed electron propagator just consists of the propagator for the bare electron field.}
   \label{fig:free_e} 
\end{figure}

\begin{figure}
\centering
   \includegraphics[scale=1]{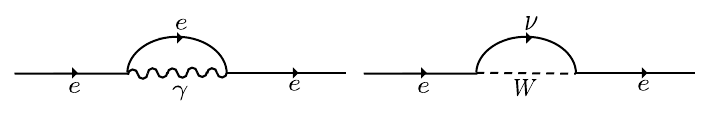}
   \caption{At higher order the physical electron gets contributions from a range of different free particles}
   \label{fig:corrections_e}
\end{figure}

Take the standard example of the self-energy of the electron (a complete calculation can be found in 18.2 of \cite{Schwartz}), the 2-point function for an electron travelling with fixed 3-momentum $\vec{p}$ is:
\begin{align}
    C(t,\vec{p})=\<\psi_{e}(p^0=\sqrt{\vec{p}^2+m_e^2},\vec{p})\bar{\psi}_e(x=0)\>.
\end{align}
At leading order, this is given by the propagator of the free electron field as shown in figure \ref{fig:free_e}.  At higher orders, the case is not so simple, the full electron propagator gets contributions from other fields in the theory, as shown in figure \ref{fig:corrections_e}. In fact, if the Weak Force is included, then there are graphs in which the electron field is replaced by the neutrino!


The problem here is \textit{not} that the physical electron contains a superposition of a range of different fields, none of the criteria given in section \ref{sec:satisfactory} rule a superposition out of a well-behaved definition of compositeness.  

The problem lies in the fact that the relative contribution from the different diagrams depends on a range of different choices.  For the most striking example of this, compare the on-shell renormalization scheme, where all higher-order diagrams get removed, to the MS scheme, where higher-order loop diagrams do contribute.  If we took these perturbative diagrams seriously when defining compositeness, we would find that the physical electron contains a W-boson - until we switch to the on-shell scheme and find that the contribution from this diagram cancels (when the electron is on-shell).


A final, but tangential, point I want to emphasize is that the physical electron (associated with a single particle state) is not a priori the same as either the electron field or the electron degree of freedom in the Lagrangian nor is the interpolating operator the same as either physical electron or the electron field.

\subsection{The full, non-perturbative argument}
In this section, I present what I consider to be the main result of this work: the full non-perturbative argument showing the incompatibility of QFT and compositeness.  An outline of the argument is given below and will be further justified throughout the rest of this section.

\syllog{P1) All S-matrix poles have the same ontological status.  That is, all S-matrix poles exist in the same way and don't carry enough information to create a hierarchy of fundamentality.} %
{P2) Two objects, $X$ and $Y$, are each individually associated with an S-matrix pole} %
{C)  The objects $X$ and $Y$ therefore have the same ontological status.  Neither is privileged over the other, the existence of one cannot depend upon or be contained within the other, and neither can be composed of the other.}

To justify the first premise, note that a pole only carries 3 pieces of information \cite{Guo_2018}. None of these provide the relevant information needed to create any form of ontological hierarchy. The location of the pole gives the mass and width of the bound state or resonance,  the residue of the pole determines the coupling strength to a given channel, and all S-matrix poles are believed to be first order and hence the order gives no extra information  (no full proof has been found but real axis poles are associated with physical states and necessarily simple and these can, under certain choices of the coupling, become resonances and all poles should have the same structure).

The second premise is merely a matter of computation. Hadron Spectroscopy calculations have confirmed the existence of S-matrix poles associated with a range of low-mass states \cite{padmanath2019hadron}.  There is, however, no reason to believe that there is (in principle) an upper mass limit to the methods of Hadron spectroscopy.  With a big enough computer, one could calculate the mass or other interesting quantity of everyday objects directly from the underlying field theory.   

Not everything that could be colloquially called an object is associated with an S-matrix pole (I go into more detail about what could constitute an object in section \ref{dennett}).  However, I do claim that any collection of matter that is somehow stuck together is associated with a pole in the S-matrix: specific atoms, molecules, chairs and people are all associated with a pole.

\subsection{Consequences of rejecting compositeness}
By rejecting compositeness, we have to accept that all bound states (or rather all poles of the S-matrix) have the same ontological status in QFT.  Electrons, photons, Pions, atoms, molecules, chairs and people are all as fundamental as each other\footnote{I have purposefully excluded quarks from this list due to asymptotic confinement - even in pure QCD they are not stable.}.

Without a well-defined notion of `made of', we cannot rigorously say `atoms are made of a proton and an electron' or that `chairs are made of atoms' or even that `an Ikea chair is made of a selection of screws and bits of wood'\footnote{although it is indeed made \textit{from} these things in the sense that by combing screws and wood results in a chair}, all of these objects are equally fundamental in QFT. 

\section{Possible objections}
In this section, I present a range of objections to the argument above and show how these objections ultimately fail to save compositeness as a well-defined notion.

\subsection{Objection 1: The usefulness of phenomenological models}

Phenomenological models based on the assumption of compositeness are widespread and it seems that every time a particle is discovered it gets categorized as a hadronic molecule, penta-quark, exotic hadron, or so on.  At the time of writing, the most recent announcement of new particles from the LHCb experiment immediately announced them as penta- and tetra-quarks \cite{SpadaroNorella:2815244,lhcbcollaboration2023amplitude,lhcbcollaboration2023observation}.  

The usefulness of compositeness in categorizing particles (not to mention the usefulness of compositeness in atomic physics, chemistry and so on) suggests that there must indeed be some underlying truth to the idea.


In response to this, firstly, none of the models based on compositeness are exact or rigorous.  The quark model can explain the quantum numbers of a particular bound state but can't be used to explain other quantities such as mass or form factor.  These models also haven't been derived directly from the underlying field theory \cite{Shepherd_2016} and are just (well-motivated) constructions.  Without the rigour that would necessarily accompany an \textit{ab inito} derivation, these models can't be taken seriously to provide a deep understanding into the nature of compositeness.  I discuss the effectiveness of approximate composite models later in section \ref{approx_comp}.

Secondly, I would also claim that such models beg the question.  There is more than sufficient freedom in constructing composite models that they cover the phenomenological landscape\footnote{especially since they only explain the quantum numbers}.  This lack of falsifiability makes it hard to view agreement with observation as evidence for compositeness.  These composite models are useful for categorization, but not for understanding the underlying physics.

\subsection{Objection 2: Quantum numbers as fundamental building blocks}

Along similar lines to composite models, we could take the quantum numbers to elucidate the constituents of a given bound state.  For example, a helium nucleus has baryon number 4 and so is made from 4 baryons, it has charge +2 so it must contain 2 positive and 2 neutral baryons - i.e. 2 protons and 2 neutrons.  It is not necessarily concerning that the constraints imposed by the set of quantum numbers don't uniquely specify the constituents, a bound state could always be a superposition of different constituents.

This would, however, lead to a notion of compositeness that is reflexive.  As an extreme example, a neutron star would be considered a composite state of many neutrons, but in reverse, a neutron would be a composite object made of a neutron star and a (slightly smaller) anti-neutron star.

It might be possible to break this reflexivity by choosing a set of particles that span all conserved quantum numbers and conclude that all bound states are composite objects of this subset of particles.  Although, I would argue that this choice is arbitrary - even if there are intuitive choices for this set of particles - defining the set would still be an external choice imposed upon the underlying field theory.

Even if a consensus on which particles get chosen to be part of this fundamental set of building blocks could be reached, it wouldn't completely save our everyday notion of compositeness.  Fully rejecting compositeness would lead to all bound states being ontologically equivalent; by instead choosing a fixed subset of fundamental particles - say we choose the particles stable in the full standard model: protons, neutrons and electrons in this set - then all bound states would be made out of these and only these.  A chair could be considered a bound state of protons, neutrons and electrons, but there would still be no satisfactory way to conclude a chair is made of atoms or molecules (or even quarks) without running into all the problems above.

\subsection{Objection 3: Bethe-Salpeter wavefunctions and the Weinberg Compositeness Criterion}
This section contains two objections, even though they are separate, the mathematics behind them is very similar and hence my response to them will be related.  I will illustrate these objections using Quantum Electrodynamics, however, none of the specifics of the theory are actually relevant.

In the $e^+e^-$ channel, the full interacting Hamiltonian $H$ contains a near-threshold\footnote{The near-threshold doesn't actually affect the overall logic, but only near-threshold bound states can be well described by perturbation theory.} bound state, Positronium.  We denote a Positronium state with fixed 3 momentum $\vec{P}$ $|B,\vec{P}\>$ (the B is for \textbf{B}ound state as $\vec{P}$ is stolen by momentum), this obeys the eigenvalue equation:
\begin{align}
    H|B,\vec{P}\>=\sqrt{M_B^2+\vec{P}^2}|B,\vec{P}\>.
\end{align}
We can use this to define a set of Bethe-Salpeter (BS) wavefunctions for the constituent particles.  If $\phi_{e^+}(x)$ and $\phi_{e^-}(x)$ are interpolating operators with the quantum numbers of a positron and electron respectively, then the momentum-space BS wavefuntion is defined as:
\begin{align}
\label{eq:BS_positronium}
    \Phi^{\vec{P}}_{e^+,e^-}(\vec{q},\vec{q}')\delta(\vec{P}-\vec{q}-\vec{q}')=\int  d^3\vec{x} d^3\vec{y}\ e^{i\vec{q}\cdot \vec{x}}e^{i\vec{q}'\cdot \vec{y}}\<0|\phi_{e^+}(t=0,\vec{x})\phi_{e^-}(t=0,\vec{y})|B,\vec{P}\>.
\end{align}
These can be constructed for any set of particles that has the combined total quantum numbers of the desired bound state. 
 These BS wavefunctions can be interpreted as encoding the momentum of the individual constituents\footnote{They are not the standard wavefunctions from NRQM as they cannot be given a probabilistic interpretation.}.  
 
 Decomposing the full Hamiltonian into a free and interacting part, $H=H_0+V$, the eigenstates of $H_0$ form a basis that is the Fock space constructed from free particles.  We can expand the bound state above in terms of these free multi-particle states:
 \begin{align}
 \label{eq:BS_expansion}
 \begin{split}
      |B,\vec{P}\>=& \int d^3 \vec{\bar{q}}\  \Phi^{\vec{P}}_{e^+e^-}(\vec{P}-\vec{\bar{q}},\vec{\bar{q}})\ |e^+(\vec{P}-\vec{\bar{q}})e^-(\vec{\bar{q}})\>_0
      \\&+\int d^3 \vec{\bar{q}} d^3\vec{\bar{q}'} \ \Phi^{\vec{P}}_{e^+e^-\gamma}(\vec{P}-\vec{\bar{q}}-\vec{\bar{q}}',\vec{\bar{q}},\vec{\bar{q}}')|e^+(\vec{P}-\vec{\bar{q}}-\vec{\bar{q}}')e^-(\vec{\bar{q}})\gamma(\vec{\bar{q}}')\>_0+\dots
 \end{split}
 \end{align}  
 Where we have included the subscript $_0$ as a reminder that the Fock states are eigenstates of the free Hamiltonian.  The BS wavefunctions are therefore the probability amplitude for the bound state to contain a certain set of free particles.  The proof of this result can be found in appendix~\ref{sec:app_BS_exp}.

 In short, the BS wavefunction tells us the behaviour of the constituent parts.  Further details of this idea can be found in refs.~\cite{hoyer_lecture,hoyer_2021,duncan2012conceptual}.

A similar objection is related to an idea from Weinberg which has come to be known as the Weinberg Compositeness Criterion - further details can be found in \cite{weinberg_elementary,weinberg_evidence}. In the above example, it is perfectly possible for the eigenstates of the free Hamiltonian to include a free positronium state $|B,\vec{P}\>_0$ - even though this is not usual for perturbative QED calculations.

He defines a quantity $Z$ as the overlap between the interacting bound-state and the equivalent free state:
\begin{align}
\label{eq:weinberg_Z}
    Z:=|\<B,\vec{P}|B,\vec{P}\>_0|^2 \quad \implies \quad 1-Z=\int_{\text{multi-particle states, }\alpha}d\alpha\ |\<B|\alpha\>_0|^2.
\end{align}
If $Z\approx 0$ then the bound state couples mainly to the multi-particle states and so is composite, and if $Z\approx 1$ then it couples strongly to the single particle state and is elementary.

For weakly bound states (such that the binding energy is small) he showed that this $Z$ can be related to the S-wave scattering length: 
\begin{align}
\label{eq:weinberg_scat_length}
    a_0\propto\frac{2(1-Z)}{2-Z}
\end{align}
making compositeness directly observable.

In summary, it seems like it is possible to expand a bound state in terms of free particle states, with this expansion containing information about the composite structure.  Furthermore, the compositeness of a bound state can, for weakly bound states, be calculated from the scattering length.

Both of these objections rely on a decomposing the Hamiltonian into free and interacting pieces.  Despite the excellent successes of perturbative QED calculations, this decomposition is often not mathematically well-defined. More rigorous treatments of scattering theory, such as that given in chapter 9 of \cite{duncan2012conceptual}, bypass this decomposition altogether and instead the "free-ness" of multi-particle states in the asymptotic past is defined in terms of transformation properties and the inner-product.

There are two responses to the objections above, firstly there is freedom in choosing the free Hamiltonian\footnote{Although not mentioned in referenced papers, Weinberg textbook \cite{weinberg_qtof} does enforce that the free Hamiltonian does have the same spectrum as the interacting Hamiltonian} which results in the constituents of a given bound state being dependent on this choice, and secondly I will show that this result is reflexive and that we can just as easily expand the electron in terms of a Fock space that includes positronium.

To show how much freedom we have when decomposing the Hamiltonian we can choose to construct the free single-particle sector out of the single-particle states of the full Hamiltonian.  Again, sticking with the positronium channel of QED, the free Hamiltonian can be constructed as\footnote{Many of the factors are a choice of normalisation, I use the choice $\<\vec{p}|\vec{p}'\>=2\omega_{\vec{p}}(2\pi)^3\delta(\vec{p}-\vec{p}')$}:
\begin{align}
    H_{0}^{single}=\int \frac{d^3\vec{p}}{2(2\pi)^3}\  \left( |e^+,\vec{p}\>\<e^+,\vec{p}|  + |e^-,\vec{p}\>\<e^-,\vec{p}| +  |B,\vec{p}\>\<B,\vec{p}|+\dots\right)
\end{align}
where  the multi-particle Fock Space is constructed from tensor products of these states.  Different Fock states are orthogonal and hence the BS wavefunction expansion given in equation \ref{eq:BS_expansion} will take the form
\begin{align}
    |B,\vec{P}\>=|B,\vec{P}\>+0\times\dots
\end{align}

This construction would automatically set Weinberg's compositeness factor in equation \ref{eq:weinberg_Z} to $Z=1$.  Using the expression for the scattering length in equation \ref{eq:weinberg_scat_length}, $Z=1$ would seem to imply that $a_0=1$. The freedom we have in $H_0$, which affects the value of $Z$, has an effect on the observed scattering length.  Although I don't go into the full derivation here, this oddity can be reconciled.  Weinberg's derivation is perturbative and makes assumptions about the relative strength of the $2\to2$ vertex relative to the $2\to1$ vertex (equation 29 of \cite{weinberg_evidence}).  The strength of these vertices is also dependent on the choice made when defining $H_0$ and therefore equation \ref{eq:weinberg_scat_length} is only valid for particular decompositions of the Hamiltonian. 

The reflexivity in these constructions comes from noting that being able to expand (fully-dressed) positronium in terms of (bare) electrons and (bare) positrons doesn't rule out the possibility of expanding the (fully-dressed) electron in terms of (bare) positronium and other states.  Alternatively, the expansion of the (fully-dressed) electron will contain (bare) positrons and then the expansion of (fully-dressed) positrons will contain (bare) electrons. The repeated use of brackets indicating which quantities are bare or fully-dressed, may seem overkill, but is needed to emphasise the difference between the two types of quantities. 

If we reject the association of bare states with physical particles the BS wavefunction expansion can't be used to provide information about constituents.  In order to view the expansion as pertaining to compositeness, we must accept some association between bare and fully-dressed states.

For the second response, the value of $Z$ for positronium doesn't necessarily constrain the value of $Z$ for the electron.  In equation 19 of \cite{weinberg_evidence}, Weinberg shows that $Z$ obeys the relation:
\begin{align}
    1-Z=\int d\alpha \frac{|\ _0\<\alpha|V|B\>|^2}{(E_{\alpha}+B)^2}
\end{align}
and therefore if the bound state is well below the relevant multi-particle threshold then $Z\approx 1$ - that is composite objects are most prevalent just below threshold.

Firstly, this doesn't enforce $Z\equiv 1$. Considering the electron as a bound state that contains Positronium, the electron would be noticeably below the multi-particle threshold however would still partly consist of positronium (just less so than positronium would consist of an electron).  Secondly, comparing the $Z$ values for different states in this way, saying that $1-Z$ scales as the inverse of binding energy, assumes the leading order 3-point vertex is constant in all cases.  This assumption does not necessarily hold over the energy scales we are discussing - after all, a major result of renormalization is the scale dependence of the coupling.

\section{An Approximate Notion of Compositeness}
\label{approx_comp}
The world looks composite of course: Ikea chairs are clearly made of planks of wood and screws and condensed matter physics has made amazing progress in explaining the properties of materials in terms of the atoms they are made of.  Individual atoms in a crystal lattice have even been photographed \cite{atom_image}.

In terms of measurable results, what matters is not association with a particular S-matrix pole, but interactions with the measuring device.  Taking the electromagnetic force as an example - it is after all the force that most directly affects our experience - the interaction depends on quantities like charge density.  If a slightly mischievous deity were to replace every electron in an Ikea chair leg with a muon - and constantly interfere with and control the motion of these muons such that they obey the same dynamics as the electrons - then the chair would look the same to any shopper that walks past.  The muonic chair would have the same interaction with the electromagnetic field as a standard chair and hence to any instrument detecting electromagnetic radiation (like our eyes), the two chairs would be indistinguishable.

Compositeness is most useful when the object is weakly bound, that is the binding energy $B$ is much less than the total rest mass $M$\footnote{We saw when discussing the Weinberg composites criterion that $Z\approx 0$ when the composite particle is near threshold.}.  I claimed earlier that compositeness is well-defined in non-relativistic quantum mechanics, which has no requirements on the interaction strength (and hence no requirements on the binding energy).  Compositeness breaks down when QFT becomes necessary, that is, the relativistic regime.  However, the expected speed of the constituents of a bound state scale with the binding energy. This can be seen classically by considering two oppositely charged objects placed infinitely far apart, as they come together and orbit each other, their speed will increase with the charge. Alternatively, in non-relativistic quantum mechanics, the expected energy of the electron in a coulomb potential scales as $Z^2 e^4$.

When $B=0$ (or negative) then the object is not bound at all and is actually two separate objects, just below the multi-particle threshold there is a sliding scale of how composite something appears.   Starting with an everyday example of screwing some Ikea chair legs together, the binding energy is on the order of maybe a couple of Joules but the rest mass is on the order of $10^{18}$J. As the binding gets stronger we enter the realm of condensed matter and chemistry, here the "atoms" can still be resolved from scattering experiments or electron microscopy \cite{kittel}.  Getting stronger, objects like hadronic molecules show some signs of compositeness but this must be otherwise inferred \cite{Guo_2018}.  Finally, the strongly bound quarks "inside" a pion are almost entirely best understood as a metaphor for categorizing the quantum numbers of the hadron.

An explicit example of this can be found in ref.~\cite{hoyer_2021}, the author shows that the Schrodinger equation for the electron-positron constituents of a positronium bound state via an expansion in terms of the momenta of the individual electrons and positrons (which scales as $p\sim \alpha m_e$ where $\alpha$ is the QED coupling).

\subsection{Can the Dennett Criterion save compositeness?}
\label{dennett}
For observational purposes, the localized charge densities in a molecule or a sheet of metal are identifiable with atoms.  The work of a condensed matter physicist will not be directly affected by this work - atoms would still be a physical part of their models.  The Dennett Criterion nicely encapsulates when something can be considered a real part of a particular model of the world (the criterion was formalised by Wallace in \cite{wallace2005everett} and was based off Dennett \cite{Dennett1991-DENRP}):

\textit{A macro-object is a pattern, and the existence of a pattern as a real thing depends on the usefulness — in particular, the explanatory power and predictive reliability — of theories which admit that pattern in their ontology.
}

\begin{figure}
    \centering
    \includegraphics[scale=0.5]{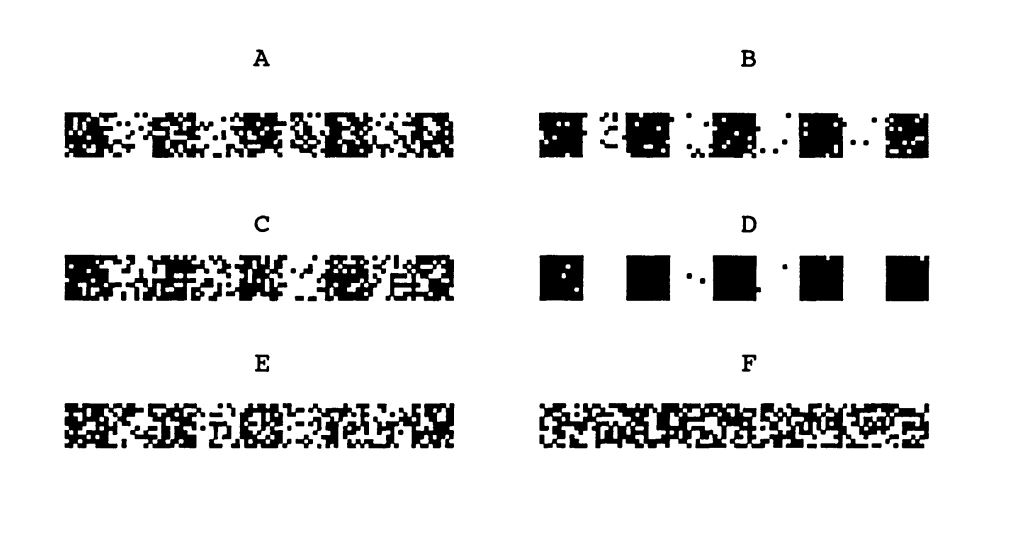}
    \caption{Six different objects, in each case the ability to describe the pattern as a checker-board varies.  Whether or not we want to take the existence of the pattern as real depends on our desired margin of error \cite{Dennett1991-DENRP}}.
    \label{fig:dennet_frames}
\end{figure}
This doesn't quite save compositeness.  Figure \ref{fig:dennet_frames} shows a series of frames made of small black and white squares.  In most of them you should be able to make out a larger checkerboard pattern, some are more obvious than others.  When developing a theory that predicts the location of the little black and white squares, it would seem a good idea to start with the larger checkerboard. According to the Dennett Criterion, the acceptance of this checkerboard pattern as real depends on the margin of error you want from the theory.  A fundamental theory (which should have no margin for error if it describes everything) will have to go deeper and will not contain this checkerboard pattern as part of its ontology.

\section{The problem of ordinary objects}

There are well-established philosophical problems that accompany our common sense understanding of what defines the term "object" - a summary of these problems can be found here \cite{sep-ordinary-objects}.  I show that, at least for some of these problems, rejecting compositeness and defining an "object" as a pole in an S-matrix\footnote{The word "object" has a much broader range of validity, but this definition creates an ontological difference between different types of objects} resolves some of these problems.  

\subsection{Problem Of The Many}
\syllog{P1) Call the chair you are sitting on "Chris".  Now consider an object consisting of all of Chris except for one particular plastic molecule, called Molly.  The new chair, with Molly removed, is called Chris Jr} %
{P2) Chris, Chris Jr and Molly all exist} %
{C)  You are sitting on (at least) 3 different objects.}

The problem here is that if an object like a chair is just a collection of atoms, then you can choose to group the constituent atoms up in any way you like - this leads to any ordinary object being spatially coincident with an extremely large number of different sets of objects.

By rejecting compositeness as a well-defined concept, it no longer makes sense to abstractly isolate and name one of the specific molecules.
\subsection{Trogs}
You are walking through a forest and you see a frog hopping merrily along by a tree.

\syllog{P1) Both the frog and the tree are just an arrangement of atoms } %
{P2) You define a new object, called a trog, which consists of the frog and the tree.  A trog is also just an arrangement of atoms} %
{C)  Trogs exist in the same way as trees and frogs.}

My initial objection is, maybe predictably, with premise 1,  frogs and trees are poles in an S-matrix and hence cannot be rigorously thought of as being an arrangement of atoms. A trog, however, is not a pole in an S-matrix and although I am not necessarily against extending the meaning of the word "object" to include entities like trogs, the fact that a trog is not an S-matrix pole means it exists differently to trees and frogs.

An alternative way of seeing this is to note that the argument above rests on the fact that trees, frogs and trogs are all just arrangements of atoms - and all arrangements of atoms exist in the same sort of way - and it is this fact that must be rejected without a rigorous definition of compositeness. 
\subsection{Material Constitution}
\syllog{P1)A piece of clay is made into a statue, both the statue and the piece of clay exist. If both of these exists then then the piece of clay is equivalent to the statue} %
{P2) Statues and pieces of clay have different properties and if they have different properties then they cannot be equivalent} %
{C)There is a contradiction between these two points.} 

In many ways this problem highlights the difficulty of relating particulars and universals - something I won't get into - however looking at this problem with the framework we have built up is insightful.  As the piece of clay gets moulded there is not a continuous process from clay to statue, instead, the piece of clay jumps from pole to pole.  Each of these jumps causes one object to stop existing and a new one to start existing.  There is still the difficulty of categorizing some of the poles as pieces of clay and some as statues - but as we saw at the start of this paper, categorizations based on phenomenology is not a good path to fundamental understanding.

\section{Conclusion}
To sum up, compositeness is not a rigorous notion within the framework of Quantum Field Theory.  Taking the mathematical structure of QFT seriously, we find that all bound states are ontologically equivalent and that different ways of trying to define the constituent parts of a bound state are either arbitrary or reflexive.  

An approximate notion of compositeness can be recovered: as the mass of a bound state approaches the multi-particle threshold, the different quantum number densities (flavour, charge etc) approach the sum of the densities of the two constituents.  As these densities determine the interaction, the bound state interacts almost as if it were a set of separate constituent parts.  This may allow compositeness in higher-level disciplines such as chemistry or solid state physics, but it doesn't recover compositeness in a rigorous or exact way.

Finally, we saw the consequences of rejecting compositeness on the philosophical problems of ordinary objects and found that many of the problems get resolved.  Although defining the term "object" to refer to a pole in an  S-matrix is maybe too restrictive, it does highlight a difference between different uses of the term.
\acknowledgments

Firstly, I want to thank my PhD supervisor, Dr Maxwell T Hansen, both for many conversations about the technical details and for his support in allowing me to pursue research outside the focus of my doctorate.

I also want to thank Dr David Wallace for the early conversations that really helped get this work off the ground and to Dr Paul Hoyer for answering many questions about his work on compositeness in gauge theories.

Finally, I want to give a special thank you to Joe Ingram for hating this idea so much that I eventually wrote a paper.
\appendix
\section{Expanding a bound state in terms of Bethe-Salpeter wavefunctions}
\label{sec:app_BS_exp}
In this appendix, we prove equation \ref{eq:BS_expansion} which shows that the fully-interacting bound state with momentum $\vec{P}$, $|B,\vec{P}\>$ can be expanded as a sum of free particle states with the BS wavefunction denoting the contribution of that specific state.  We only calculate the details for the $e^+e^-$ terms but the calculation generalizes nicely.  All interpolation operators will be inserted at equal time, which we will set to be $t=0$.

The most general expansion of the bound state is of the form:
\begin{align}
    |B,\vec{P}\>=\int d^3\vec{\bar{q}}\  f(\vec{P}-\vec{\bar{q}},\vec{\bar{q}}) |e^+(\vec{P}-\vec{\bar{q}}),e^-(\vec{\bar{q}})\>_0+\text{other particle content}
\end{align}
where $f$ is some arbitrary function of the momenta of the two particles.  Momentum conservation requires that the total momenta of the two-particle state is equal to the momenta of the bound state.

Therefore, the inner product with a free state of two particles with arbitrary momentum is:
\begin{align}
    _0\<e^+(\vec{q})e^-(\vec{q}')|B,\vec{P}\>=&\int d^3\vec{\bar{ q}}\ f(\vec{P}- \vec{\bar{q}}, \vec{\bar{q}})\ _0\<\phi_+(\vec{q})\phi_-(\vec{q}')|e^+(\vec{P}-\vec{\bar{q}}),e^-(\vec{\bar{q}})\>_0\\
    =&\int  d^3\vec{\bar{q}}\ f(\vec{P}-\vec{q},\vec{q}) (2\pi)^6\delta(\vec{P}-\vec{\bar{q}}-\vec{q})\delta(\vec{\bar{q}}-\vec{q}')\\
    \label{eq:f_to_compare}
    =&f(\vec{q},\vec{q}')(2\pi)^6\delta(\vec{P}-\vec{q}-\vec{q}').
\end{align}
We have chosen a non-relativistic normalization of states in order to simplify the presentation.  We now need to relate $f$ to the BS wavefunction.

Starting with the definition of the momentum-space BS-wavefunction as given in equation \ref{eq:BS_positronium}:
\begin{align}
    \Phi^{\vec{P}}_{e^+,e^-}(\vec{q},\vec{q}')\delta(\vec{P}-\vec{q}-\vec{q}')=\int  d^3\vec{x} d^3\vec{y}\ e^{i\vec{q}\cdot \vec{x}}e^{i\vec{q}'\cdot \vec{y}}\<0|\phi_{e^+}(\vec{x})\phi_{e^-}(\vec{y})|B,\vec{P}\>
\end{align}
This is equivalent to the definition given in \cite{duncan2012conceptual} except we have Fourier transformed to momentum space and included an extra delta to account for having two free coordinates\footnote{In the reference, they pick a coordinate system where the particles are at $\pm x/2$, or equivalently they both have equal momenta}.  The RHS has an implicit delta function coming from translation invariance, and hence this is included on the left as well.

The interpolation operators are inserted at equal time, which we set to $t=0$, as the energy is determined by the particle content of the state.  
Insert a complete set of non-interacting states between the interpolation operators and the bound state.  The only non-interacting states that couple to the interpolation operators include an $e^+,\ e^-$ pair
\begin{align}
\begin{split}
        \Phi^{\vec{P}}_{e^+,e^-}(\vec{q},\vec{q}')\delta(\vec{P}-\vec{q}-\vec{q}')&=\int \frac{d^3\vec{\bar{q}}}{(2\pi)^3}\frac{d^3\vec{\bar{q}'}}{(2\pi)^3}\int  d^3\vec{x} d^3\vec{y}\ e^{i\vec{q}\cdot \vec{x}}e^{i\vec{q}'\cdot \vec{y}}\\& \quad \times\<0|\phi_{e^+}(\vec{x})\phi_{e^-}(\vec{y})|e^+(\bar{q})e^-(\bar{q}')\>_0\ _0\<e^+(\bar{q})e^-(\bar{q}')|B,\vec{P}\>.
\end{split}
\end{align}
As the states are free we have the following result
\begin{align}
   \<0|\phi_{e^+}(\vec{x})\phi_{e^-}(\vec{y})|e^+(\bar{q})e^-(\bar{q}')\>_0=e^{-i\vec{x}\cdot\bar{\vec{q}}}e^{-i\vec{y}\cdot\bar{\vec{q}}'}
\end{align}
where the interpolation operators are assumed to have unit normalization.
This becomes
\begin{align}
\begin{split}
        \Phi^{\vec{P}}_{e^+,e^-}(\vec{q},\vec{q}')\delta(\vec{P}-\vec{q}-\vec{q}')&=\int \frac{d^3\vec{\bar{q}}}{(2\pi)^3}\frac{d^3\vec{\bar{q}'}}{(2\pi)^3}\int  d^3\vec{x} d^3\vec{y}\ e^{i(\vec{q}-\bar{\vec{q}})\cdot \vec{x}}e^{i(\vec{q}'-\bar{\vec{q}}')\cdot \vec{y}}\\& \qquad \qquad\times \ _0\<e^+(\bar{\vec{q}})e^-(\bar{\vec{q}}')|B,\vec{P}\>
\end{split}\\
&=\ _0\<e^+(\vec{q})e^-(\vec{q}')|B,\vec{P}\>\\
\bigg(&=\delta(\vec{P}-\vec{q}-\vec{q'})\ _0\<e^+(\vec{q})e^-(\vec{P}-\vec{q})|B,\vec{P}\>\bigg).
\end{align}

Comparing this to equation \ref{eq:f_to_compare} we find that:
\begin{align}
    f(\vec{q},\vec{q}')= \Phi^{\vec{P}}_{e^+,e^-}(\vec{q},\vec{q}')
\end{align}

and thus the bound state can be written as an expansion in terms of free particle states where the coefficient is proportional to the BS wavefunction.

\newpage

\theendnotes

\bibliographystyle{JHEP} 
\bibliography{refs.bib}
\end{document}